%% file: short_paper.tex
\documentclass[a4paper,UKenglish,cleveref,autoref,thm-restate]{lipics-v2021} 
\pdfoutput=1

\title{Analyzing the Availability of E-Mail Addresses for PyPI Libraries}

\titlerunning{Analyzing the Availability of E-Mail Addresses for PyPI Libraries}

\author{Alexandros Tsakpinis}
{fortiss GmbH, Munich, Germany}
{tsakpinis@fortiss.org}
{https://orcid.org/0000-0001-6561-2866}
{}

\author{Alexander Pretschner}
{Technical University of Munich, Munich, Germany}
{alexander.pretschner@tum.de}
{https://orcid.org/0000-0002-5573-1201}
{}

\authorrunning{A. Tsakpinis and A. Pretschner}

\Copyright{Alexandros Tsakpinis and Alexander Pretschner}



\relatedversion{}

\acknowledgements{}

\hideLIPIcs 

\begin{document}
\nolinenumbers
\maketitle

\input{sections/0_Abstract}
\input{sections/1_Introduction}
\input{sections/2_Background_and_Related_Work}
\input{sections/3_Research_Method}
\input{sections/4_Results}
\input{sections/5_Discussion}
\input{sections/6_Threats_to_Validity}
\input{sections/7_Conclusion_and_Future_Work}

\input{sections/8_Data_Availability}

\bibliography{short_paper}

\end{document}

%% file: sections/0_Abstract.tex
\begin{abstract}
    \textbf{Background:}
    Open Source Software (OSS) libraries form the backbone of modern software systems, yet their long-term sustainability often depends on maintainers being reachable for support, coordination, and security reporting. 
    \textbf{Aims:}
    In this paper, we empirically analyze the availability of contact information, specifically e-mail addresses, across 754,413 Python libraries on the Python Package Index (PyPI) and their associated GitHub repositories. 
    \textbf{Method:}
    We examine where maintainers provide this information, assess its validity, and explore coverage across individual libraries and their dependency chains. 
    \textbf{Results:}
    Our findings show that 79.1\% of libraries include at least one valid e-mail address, with PyPI serving as the primary source (76.5\%). When analyzing dependency chains, we observe that up to 97.7\% of direct and 97.5\% of transitive dependencies provide valid contact information. At the same time, we identify over 793,000 invalid entries, primarily due to missing fields. 
    \textbf{Conclusions:}
    Our results indicate strong maintainer reachability, while highlighting opportunities for improvement, such as offering clearer guidance to maintainers during the packaging process and introducing opt-in validation mechanisms for existing e-mail~addresses.
\end{abstract}

%% file: sections/1_Introduction.tex
\section{Introduction}
Over the past decades, Open Source Software (OSS) has evolved into a fundamental element of the software product life cycle, supported by commercial, technical, and quality reasons~\cite{ebert2008open}. Today, OSS components constitute the vast majority of code in commercial systems, with estimates ranging between 80\% and 90\%, illustrating the degree of reliance the industry places on them~\cite{oss2022, pittenger2016open}. Among the most influential OSS artifacts are software libraries (often distributed as packages), which have become indispensable building blocks in modern development. By reusing proven and well-tested code to implement specialized functionalities, developers can avoid redundant work and speed up development cycles~\cite{bauer2012structured}. Once incorporated into a project a library is treated as a dependency~\cite{cox2019surviving}, usually resulting in a dependency chain in which each dependency is itself reliant on other dependencies. This interconnection introduces significant risks. OSS dependencies may become compromised through vulnerabilities (e.g., Log4j~\cite{log4j2021}) or supply chain attacks (e.g., LiteLLM~\cite{litellm}, XZ Utils~\cite{xz2024}), creating widespread challenges for the ecosystem~\cite{decan2018impact}. Although OSS communities typically act quickly to publish patched versions when vulnerabilities are discovered~\cite{rahkema2022swiftdependencychecker}, long-term sustainability is not guaranteed. Libraries may lose support entirely, or maintenance may be suspended, leaving users with unresolved issues~\cite{bauer2012structured, raemaekers2011exploring}. Because OSS libraries are highly interconnected, the consequences of an abandoned or unmaintained dependency can ripple through the ecosystem, affecting thousands of projects through direct and transitive dependencies~\cite{kula2014visualizing, tsakpinis2023analyzing}.

While academia and industry emphasize the importance of sustaining OSS maintenance through financial and non-financial support~\cite{linaaker2024sustaining, medappa2023sponsorship, tidelift2024}, identifying the right forms of support for individual maintainers remains challenging. Maintainers may communicate support-related topics publicly via GitHub issues or discussions and receive financial contributions through platforms such as GitHub Sponsors, for which prior work has examined the accessibility of these channels~\cite{tsakpinis2024analyzing, tsakpinis2025analyzing}. However, sensitive and security-related matters are often better addressed through private communication channels such as e-mail, whose availability in the PyPI ecosystem remains largely unexplored. This gap is critical, as missing contact details can hinder external stakeholders from providing support or coordinating on sensitive issues. To address this gap, we analyze the PyPI ecosystem from two perspectives: (1) analyzing libraries individually without considering their dependencies, and (2) including their dependency chains, which capture cascading effects across the ecosystem. The following research questions~(RQs) are designed to gather evidence and provide these insights:

\textbf{RQ1:} What is the distribution of sources where e-mail addresses are provided?

\textbf{RQ2:} What is the distribution of invalid e-mail addresses across the PyPI ecosystem?

\textbf{RQ3:} What is the ratio of PyPI libraries providing e-mail addresses?
 
\textbf{RQ4:} What is the ratio of PyPI libraries in dependency chains providing e-mail addresses?


We focus on the PyPI ecosystem in our study, as Python is widely adopted in modern software development and characterized by its strong reliance on library reuse and integration~\cite{abdalkareem2020impact, decan2016topology, Octoverse2025}. For each library available on PyPI, we extract declared e-mail addresses and, where available, complement them with additional information from GitHub repositories, such as repository owner type (individual or organization) and e-mail addresses listed there. GitHub is chosen as the primary code management platform, given its dominant role in hosting OSS projects, particularly those distributed through PyPI~\cite{eghbal2020working, tsakpinis2024analyzing}. Because OSS libraries rarely exist in isolation, evaluating them individually offers only a partial picture. We therefore extend our analysis to their dependency structures. For every library, we retrieve its direct dependencies and construct a dependency graph, augmented with available e-mail address information. To account for the varying degrees of influence libraries hold within the ecosystem, we calculate their relative importance using PageRank scores derived from the dependency graph~\cite{mujahid2021toward, tsakpinis2024analyzing, tsakpinis2025analyzing}. This graph-based representation provides the analytical foundation to assess the availability of contact information for individual libraries and their dependency chains across subsets of libraries with varying ecosystem importance.

Our key contributions consist of several empirical insights into the availability of contact information in the PyPI ecosystem. First, we observe that the majority of libraries (57.3\%) include e-mail addresses exclusively on their PyPI project page, while only a small fraction (2.6\%) rely solely on GitHub. An additional 19.2\% of libraries list addresses on both platforms, whereas 20.9\% offer no contact details at all. These results underscore PyPI's role as the dominant channel for maintainers to share contact information. Second, we identify 793,780 invalid e-mail entries across both platforms, with the majority (77.7\%) originating from GitHub. Most invalid entries stem from empty fields (95.1\%), but we also observe undeliverable (4.3\%) and syntactically incorrect (0.6\%) addresses. These issues call for two distinct measures: clearer guidance during package submission to encourage maintainers to provide contact information in the first place, and automated validation---offered as an opt-in feature---to detect formatting and delivery problems. Third, our analysis shows that 79.1\% of libraries include at least one valid e-mail address, either on PyPI or GitHub, with this number rising to 95.8\% among the top 0.1\% of libraries as measured by PageRank. Lastly, we analyze the availability of e-mail addresses in dependency chains: up to 97.7\% of direct and up to 97.5\% of transitive dependencies provide valid e-mail addresses. In contrast, libraries not used as dependencies show lower coverage, indicating an opportunity to improve metadata completeness. These findings offer a positive signal for developers relying on PyPI libraries, indicating that most libraries and especially their dependencies, which are integrated automatically without direct control, remain reachable through valid e-mails.

%% file: sections/2_Background_and_Related_Work.tex
\section{Background and Related Work}

A wide range of work has examined dependency networks in software ecosystems. Prior studies investigate security concerns such as vulnerabilities~\cite{alfadel2023empirical, dusing2022analyzing}, the injection of malicious code~\cite{guo2023empirical}, and the robustness of ecosystems over time~\cite{hafner2021node}. Other research considers package versioning and update practices~\cite{javan2023dependency}, challenges of configuration management~\cite{peng2023less}, dependency conflicts~\cite{wang2020watchman}, code smells~\cite{cao2022towards} and errors~\cite{mukherjee2021fixing}. Further perspectives include the analysis of widely used NPM packages~\cite{mujahid2023characteristics}, the sustainability of OSS Python projects~\cite{valiev2018ecosystem}, and the prevalence of trivial packages in PyPI and NPM~\cite{abdalkareem2020impact}. While these studies provide valuable insights into the dynamics of ecosystems and their libraries, they do not examine how maintainers provide contact information.
In addition to research on individual ecosystem aspects, other work has emphasized structural characteristics, especially dependencies~\cite{bommarito2019empirical, decan2016topology}, particularly focusing on transitive dependencies~\cite{decan2019empirical, kikas2017structure, tsakpinis2024analyzing, tsakpinis2025analyzing}. Despite the importance of these structural features, the availability of contact information within dependency chains has not been investigated, leaving an important gap.
Complementing these ecosystem-focused studies, a small body of work has directly considered contact information of maintainers. Zahan et al. showed that expired e-mail domains in npm packages represent weak links that attackers could exploit for supply-chain compromises~\cite{zahan2021weaklinks}. Others demonstrated that GitHub commit metadata, including e-mail addresses, can be harvested for phishing attacks~\cite{knothe2019metadata}. Alarcon et al. explored  how metadata influences trust in OSS, noting that visible maintainer contact details can shape adoption decisions~\cite{gregg2025metadataTrust}. These studies illustrate the risks and the social value of contact information, but none have assessed the availability of valid e-mail addresses across PyPI packages and their dependency chains, which is the focus of this work.

%% file: sections/3_Research_Method.tex
\section{Research Method}
In this empirical study, we use large-scale repository mining and quantitative analysis to investigate the availability of e-mail addresses in PyPI libraries.

\subsection{Data Collection}
\label{sec:data_collection}
To examine the availability of e-mail addresses in the PyPI ecosystem, we collected library metadata from multiple sources using custom Python scripts. As a starting point, a comprehensive list of all available libraries was retrieved via an endpoint~\cite{pypi-simple}. For each library, more detailed information including declared e-mail addresses and dependency data was obtained from a second PyPI endpoint~\cite{pypi-json}. We extracted all direct and optional dependencies for the latest version of each library and constructed a dependency graph with unrestricted transitive depth. For projects linked to GitHub, we further relied on the GitHub API to determine whether the repository owner was an individual or an organization, and to gather e-mail addresses listed on the repository owner’s profile. In addition, we inspected the presence of a \texttt{SECURITY.md} file in the repository, treating it as an equally valid private communication channel, as it typically specifies how maintainers can be contacted for sensitive or non-public matters~\cite{kanaji2025empirical}. Each e-mail address collected from PyPI and GitHub is further validated using an external library to check for syntactic correctness and domain resolvability, which serves as proxy for potential deliverability~\cite{email_validator_pypi}. The resulting dataset was stored in JSON format, mapping each library to data retrieved from PyPI and additional information from GitHub.

\subsection{Data Analysis}
While analyzing the collected data, we adopt standard terminology such as nodes, dependency chains, and dependency graph. Each library is modeled as a node, with directed edges representing its direct dependencies. Together, these connections form a dependency graph that captures the structure of the PyPI ecosystem. The dataset described in Section~\ref{sec:data_collection} provides the foundation for two complementary perspectives. First, we treat each library as an independent node, ignoring its dependencies. This library-centric perspective offers a straightforward view of the ecosystem, focusing on libraries in isolation. Second, we expand the analysis to dependency chains, which trace the paths from a library through its direct and transitive dependencies. This perspective uncovers structural patterns and cascading effects that cannot be observed when examining libraries individually. To address our research questions, we compute four metrics: (1) the distribution of sources where e-mail addresses are provided, (2) the distribution of reasons for invalid e-mail addresses, (3) the ratio of libraries providing e-mail addresses, (4) and the ratio of libraries in dependency chains providing e-mail addresses.
The third and fourth metrics are further analyzed across different subsets of the dependency graph. To identify these subsets, we apply the PageRank algorithm, which assigns an importance score to each node based on its incoming and outgoing dependencies, following approaches from related work~\cite{mujahid2021toward, tsakpinis2024analyzing, tsakpinis2025analyzing}. Categorizing libraries by their relative importance enables a more fine-grained analysis, highlighting differences in the availability of e-mail addresses across varying levels of ecosystem importance.

%% file: sections/4_Results.tex
\section{Results}

Data collection started on April 7, 2026, and spanned approximately three days, constrained by GitHub’s API rate limits. This process resulted in an initial dataset containing 778,976 PyPI libraries. During dependency parsing, libraries were excluded that were not part of the initial ecosystem library list~\cite{pypi-simple}. This step led to the removal of 1.5\% of libraries from a node and 1.7\% from a dependency perspective. Furthermore, 3.2\% of libraries were excluded due to errors encountered while querying the detailed ecosystem endpoint~\cite{pypi-json}, primarily caused by 404 status codes. These errors typically occur when package names retrieved from the comprehensive library list endpoint~\cite{pypi-simple} could not be resolved through the detailed endpoint~\cite{pypi-json}. After these filtering steps, the final dataset comprised 754,413 PyPI libraries connected through 2,133,548 dependencies, sourced from 97,175 unique libraries.

The subsequent analysis focuses on libraries with increasing importance, as measured by their PageRank scores. Libraries with higher PageRank values have greater influence within the dependency graph and are more frequently depended upon. In Figures \ref{fig:results_node_perspective} and \ref{fig:percentage_nodes_having_email}, the x-axis represents subsets of the dataset used for calculation. For instance, $x=10$ refers to the top 10\% of libraries ranked by PageRank, whereas $x=100$ includes the entire dataset, encompassing both high- and low-importance libraries. Unless otherwise specified, the terms contact information and e-mail address are used interchangeably in the upcoming sections.

\subsection{RQ1+RQ2+RQ3: Library Perspective}
\label{sec:node_perspective}

\begin{figure}[ht]
    \centering
    \includegraphics[width=0.91\textwidth, trim=0 2mm 0 2mm, clip]{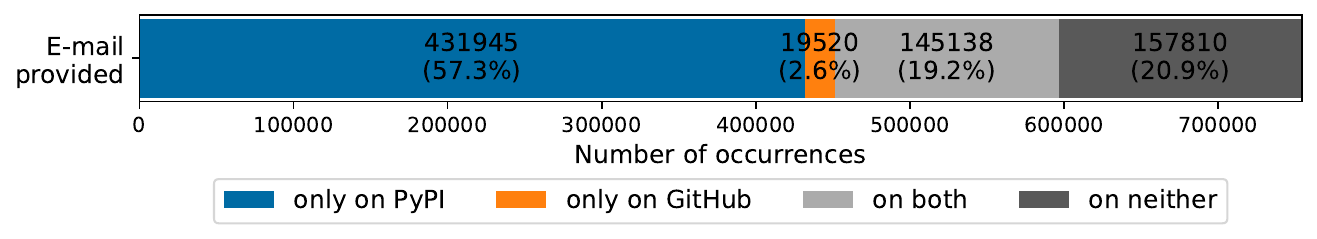}
    \caption{Distribution of sources for e-mail addresses}
    \label{fig:results_distribution_email_sources}
\end{figure}

\begin{figure}[ht]
    \centering
    \includegraphics[width=0.91\textwidth, trim=0 2mm 0 2mm, clip]{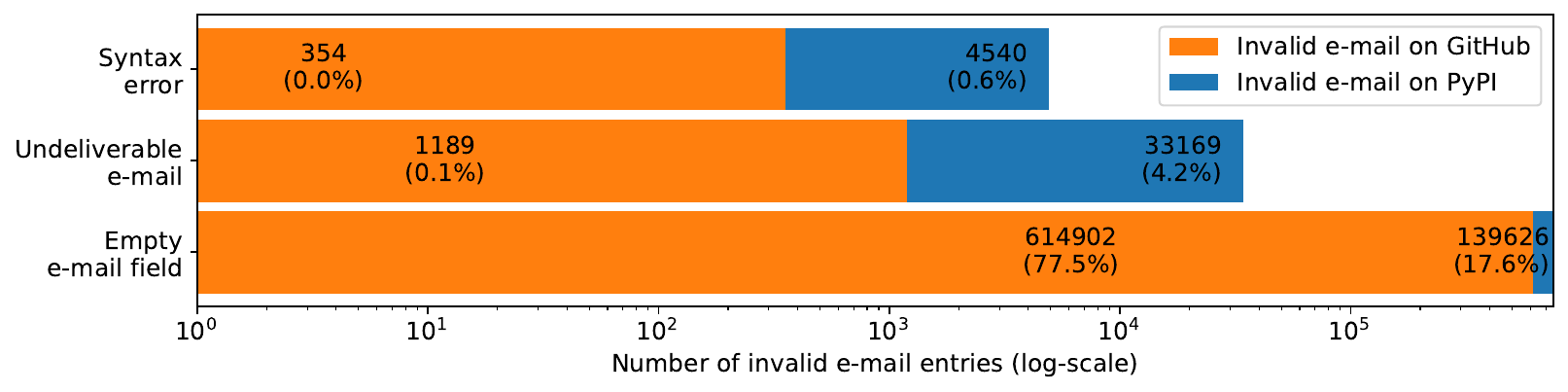}
    \caption{Distribution of reasons for invalid e-mail addresses}
    \label{fig:reasons_for_invalid_emails}
\end{figure}

\begin{figure}[ht]
    \centering
    \includegraphics[width=0.82\textwidth, trim=0 2mm 0 2mm, clip]{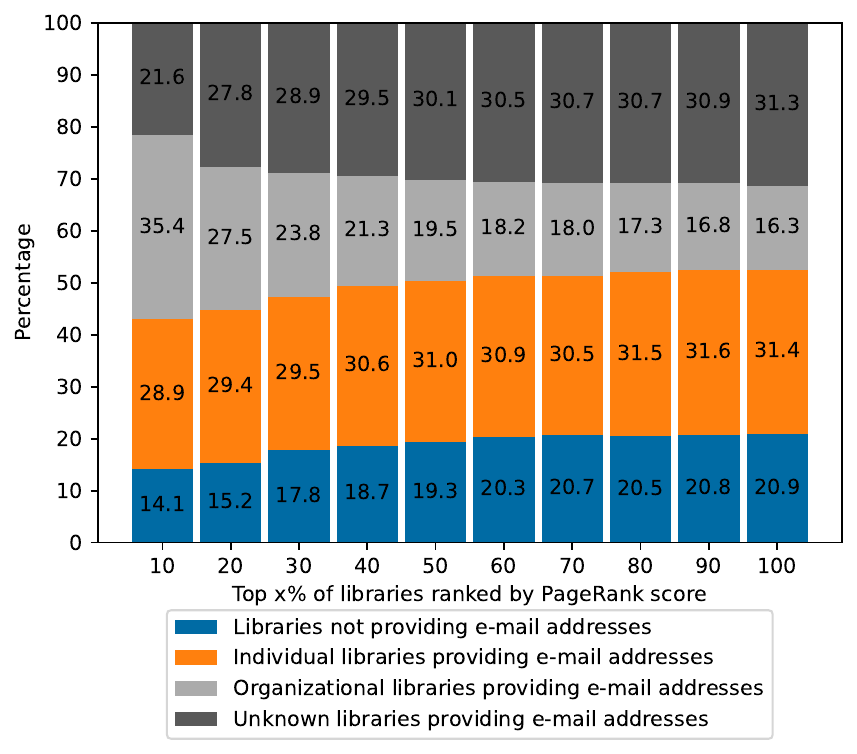}
    \caption{Ratio of PyPI libraries providing e-mail addresses}
    \label{fig:results_node_perspective}
\end{figure}

Figures \ref{fig:results_distribution_email_sources}, \ref{fig:reasons_for_invalid_emails}, and \ref{fig:results_node_perspective} present findings related to the characteristics of individual nodes within the dependency graph, without taking their dependency chains into account.

Figure \ref{fig:results_distribution_email_sources} illustrates the distribution of sources where e-mail addresses are provided across the analyzed libraries. The majority of libraries (57.3\%) provide contact information exclusively on PyPI. In contrast, only 2.6\% list e-mail addresses solely on GitHub. An additional 19.2\% of libraries offer contact information on both platforms. Finally, 20.9\% of libraries do not provide any e-mail address on either PyPI or GitHub. These results underline PyPI's central role as the primary source of contact information within the ecosystem, while e-mail addresses are oftentimes omitted on GitHub.

To investigate the reasons for invalid e-mail addresses on PyPI and GitHub, Figure \ref{fig:reasons_for_invalid_emails} breaks down libraries with empty e-mail fields, syntactically incorrect addresses, or undeliverable e-mail entries, focusing on all libraries that do NOT provide e-mail addresses on both platforms (light gray category in Figure \ref{fig:results_distribution_email_sources}). Here, undeliverable means that the e-mail address references an unresolvable domain, suggesting that delivery attempts would likely fail. 

In total, 793,780 invalid e-mail entries were identified, with 616,445 (77.7\%) originating from GitHub and 177,335 (22.3\%) from PyPI. Note that the total number of invalid e-mail entries exceeds the number of distinct libraries, as a single library may be associated with multiple URLs on PyPI, or may exhibit issues on both platforms resulting in duplicate counts. The most common issue was empty e-mail fields, which accounted for 754,528 entries (95.1\%) overall---614,902 (77.5\%) from GitHub and 139,626 (17.6\%) from PyPI. Undeliverable e-mail addresses were the second most frequent issue, totaling 34,358 entries (4.3\%), including 33,169 (4.2\%) from PyPI and 1,189 (0.1\%) from GitHub. The least common category was syntax errors, contributing 4,894 entries (0.6\%)---4,540 (0.6\%) from PyPI and 354 (0.045\%) from GitHub. These results suggest that while PyPI occasionally includes malformed or undeliverable e-mail addresses, GitHub more frequently omits contact information entirely, making it a less reliable source for valid e-mail data. Although syntax and deliverability issues occur relatively rarely, PyPI could consider offering automated checks as an opt-in feature for maintainers, similar to suggestions in related work for verifying assigned URLs~\cite{tsakpinis2024analyzing,tsakpinis2025analyzing}, to provide feedback on whether the listed contact information remains reachable.

After identifying the general distribution of e-mail sources and the reasons for invalid e-mail addresses, Figure \ref{fig:results_node_perspective} presents a breakdown of libraries that provide e-mail addresses, grouped by maintainer type. Considering the full dataset ($x=100$), 79.1\% of libraries provide an e-mail address, while 20.9\% do not. Among those that include an e-mail address, 31.4\% are maintained by individuals, 16.3\% by organizations, and 31.3\% are libraries for which the owner type could not be determined because they lack an assigned repository.

Focusing on more important libraries by narrowing the scope from $x=100$ to $x=10$, the share of libraries offering contact information increases. For the top 10\% of libraries ranked by PageRank, 85.9\% provide e-mail addresses, while 14.1\% do not. Within this subset, 35.4\% are maintained by organizations, 28.9\% by individuals, and 21.6\% fall into the unknown category. This trend suggests that high-impact libraries are more likely to offer contact information. Interestingly, the share of organizationally developed libraries more than doubles compared to the full dataset, while the proportion of individually developed libraries decreases slightly. Additional analysis, not shown in the figure, confirms that this pattern persists for the top 0.1\% (754) of libraries. In this group, 95.8\% (722) provide an e-mail address---27.1\% from individuals, 59.8\% from organizations, and 8.9\% from unknown sources. These results further emphasize the upward trend among organizationally maintained libraries and the decline among individually maintained ones. Among these top 0.1\% of libraries, 4.2\% (32) remain without any contact information. A manual examination of these cases indicates that contact to the maintainers could still be established via links to their GitHub repositories (24) or to their project homepage (6). One library deliberately removed its contact details while pointing to a successor package, and for the last case no contact information could be identified at all. Although such workarounds may be feasible, the manual effort required prevents the integration of maintainer contact discovery into automated workflows.

Another observation is that more important libraries are not only more likely to include e-mail addresses but also more frequently link to their GitHub repositories. This is reflected in the continuous decline of the last category ("\textit{Unknown libraries providing e-mail addresses}"). This trend is promising for related work~\cite{tsakpinis2024analyzing}, as it suggests that libraries that prioritize providing contact information also enable automated monitoring of their maintenance activities on GitHub through frameworks such as the OpenSSF Scorecard~\cite{openssf_scorecard}.

\subsection{RQ4: Dependency Chain Perspective}
\label{sec:dependency_chain_perspective}

\begin{figure}[ht]
    \centering
    \includegraphics[width=0.82\textwidth, trim=0 2mm 0 2mm, clip]{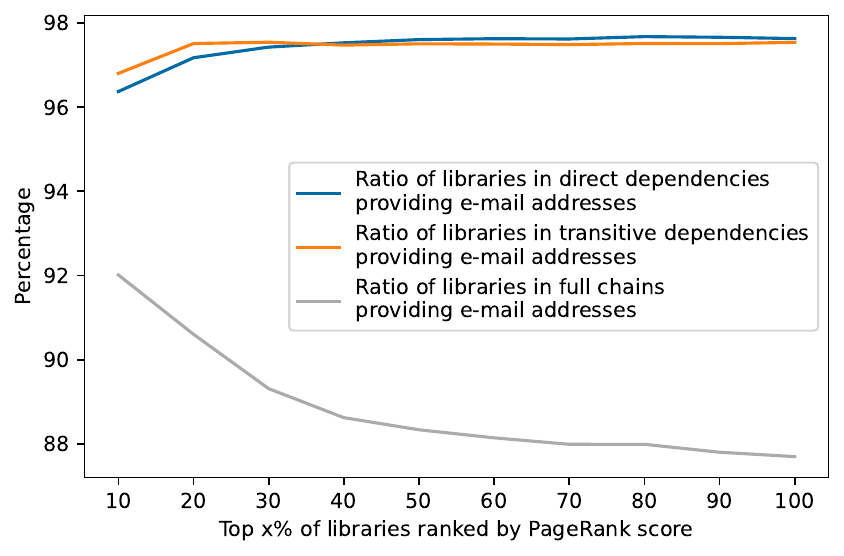}
    \caption{Ratio of PyPI libraries in dependency chains providing e-mail addresses}
    \label{fig:percentage_nodes_having_email}
\end{figure}

Figure \ref{fig:percentage_nodes_having_email} presents the results of the analysis on the availability of e-mail addresses across libraries and their dependency chains, evaluated from three distinct perspectives. The analysis considers: (1)~each PyPI library together with its full set of direct and transitive dependencies, referred to as the \textit{full chain}; (2) only the direct dependencies of each library; and (3) only the transitive dependencies. This separation is important, as direct dependencies are explicitly selected and managed by library maintainers, while transitive ones are included implicitly and lie outside of direct control. For each of these perspectives, the results are aggregated across all libraries in the ecosystem to offer a holistic view of contact information availability within the PyPI dependency network.

Figure \ref{fig:percentage_nodes_having_email} shows that as library importance increases, the share of libraries within full dependency chains that provide an e-mail address rises from 87.7\% across the full dataset to 92.0\% among the top 10\% of libraries. Even higher ratios are observed when focusing solely on direct (96.4\% to 97.7\%) and transitive dependencies (96.8\% to 97.5\%). The relatively narrow range of these values suggests a stable set of libraries underlying both dependency types, consistent with the fact that dependencies are drawn from a comparatively smaller pool of 97,175 unique libraries, compared to 754,413 libraries in total, which are reused across the dependency chains. This consistent gap between full chains and their direct and transitive dependencies indicates that libraries not used as dependencies, namely those positioned at the entry point of a dependency chain, are more likely to omit contact information, whereas libraries used as dependencies tend to include e-mail addresses more reliably. Since maintainers have full control over their own package metadata, this gap could be addressed by encouraging them to include contact information during package publication. At the same time, the high share of libraries with e-mail addresses in direct and transitive dependencies may serve as a positive signal for developers relying on the PyPI ecosystem. It indicates that most dependencies, which are integrated automatically and without direct control, provide valid contact details. This suggests a general awareness of maintenance responsibilities and a commitment to communication within the ecosystem.

%% file: sections/5_Discussion.tex
\section{Discussion}

This study provides empirical insights from a large-scale analysis on the availability of contact information, focusing on e-mail addresses in the PyPI ecosystem and its dependency structures. The overall availability of e-mail addresses (79.1\%) indicates that many maintainers make contact information available, with higher availability among more central libraries. This effect is even stronger when considering dependency structures: although root libraries may lack contact information, their direct and transitive dependencies provide it at very high rates (up to 97.7\% and 97.5\%), indicating that the ecosystem as a whole remains largely reachable. For practitioners, this suggests that critical upstream components are associated with maintainers for whom at least one contact channel is available on PyPI or GitHub.

At the same time, it is important to revisit the assumption that maintainers universally aim to be reachable via e-mail. Prior work has shown that publicly exposed e-mail addresses can be exploited~\cite{knothe2019metadata}. In response to such concerns, GitHub has introduced mechanisms to protect maintainer privacy, including masking e-mail addresses in commits and restricting their retrieval via APIs~\cite{blocking_email}. From this perspective, the substantial share of libraries missing e-mail addresses on GitHub may not solely indicate incomplete metadata, but can also reflect deliberate decisions by maintainers to limit unsolicited or potentially harmful communication.

In line with this, PyPI emerges as the primary source of e-mail addresses, whereas GitHub shows a higher proportion of missing entries. This divergence may indicate differing expectations between platforms: while PyPI encourages structured metadata provision during package publication, GitHub offers more flexibility and stronger privacy controls, which may lead maintainers to withhold direct contact information. Consequently, improving reachability should not be interpreted as universally increasing the availability of e-mail addresses, but rather as balancing accessibility with privacy considerations.

In this context, alternative communication channels become particularly relevant. Dedicated security contact files (e.g., \texttt{SECURITY.md}) can provide more controlled and context-specific means of interaction, especially for sensitive topics such as vulnerability disclosure. At the same time, our findings suggest that such channels currently play a complementary rather than dominant role. Among the libraries for which contact information is provided on GitHub, only 22.5\% include a \texttt{SECURITY.md} file. This indicates that contact information on GitHub is still more commonly provided through e-mail than through alternatives such as dedicated security contact files. Encouraging the adoption of such channels may therefore represent a useful way to support maintainers who prefer more controlled communication~mechanisms.

Lightweight interventions, such as nudges during package publication or optional validation mechanisms, could still help improve metadata completeness for maintainers who are willing to provide contact information. In particular, nudges may address the high number of empty fields by encouraging maintainers to provide contact details, while validation mechanisms could help identify syntactically incorrect or undeliverable e-mail addresses. These improvements would primarily support the remaining fraction lacking valid contact details, without imposing requirements on maintainers who prefer not to disclose them.

Overall, our findings suggest that the PyPI ecosystem demonstrates strong maintainer reachability, particularly among influential components, while also highlighting the need to consider privacy, security, and maintainer preferences when interpreting the absence of contact information. Rather than treating missing e-mail addresses solely as a deficiency, they should be understood as part of a broader trade-off between openness and protection, with implications for how communication channels are designed in open-source ecosystems.

%% file: sections/6_Threats_to_Validity.tex
\section{Threats to Validity}
We discuss potential threats to validity along the following four dimensions~\cite{runeson2009guidelines}:

\textbf{Construct validity:}
A minor threat arises from the assumption that the presence of an e-mail address implies that the maintainer is reachable. To reduce this risk, we only include addresses verified as deliverable using the \texttt{validate\_email} function of the \texttt{email\_validator} package~\cite{email_validator_pypi}. Complementary evidence from a large-scale survey study targeting 50,000 e-mail addresses on PyPI shows that only about 11\% of contacted addresses were undeliverable, suggesting that the majority of e-mail addresses are indeed reachable in practice~\cite{tsakpinis2026authors}. However, deliverability does not guarantee that maintainers actively monitor or respond to these addresses. To further assess this concern, we conducted an additional analysis of the top 0.1\% most important libraries ranked by PageRank. The results show that the vast majority of e-mails are associated with established domains (e.g., \texttt{gmail.com} or other organization-specific domains), while only a negligible fraction corresponds to \texttt{noreply} addresses (0\% on GitHub and 0.7\% on PyPI), which typically indicate addresses not intended for direct communication. This suggests that most provided e-mails are intended for actual communication, partially mitigating this threat. Taken together, the verification of deliverability and the prevalence of established e-mail providers, along with complementary evidence from prior survey-based work, offer supporting evidence that e-mail availability constitutes a reasonable, though imperfect, proxy for maintainer reachability. Another threat concerns dependency resolution: version information was ignored, and only the latest version of each dependency was considered. While this simplification may not fully reflect the ecosystem’s dynamics, we expect effects on the PageRank scores to be negligible.

\textbf{Internal validity:} 
Errors may occur during data collection, for example due to missing dependencies or overlooked e-mails. To mitigate this risk, we repeated the data collection process for failed entries and manually spot-checked random samples.

\textbf{External validity:} 
Our study is limited to the PyPI ecosystem and GitHub as the code management system (CMS). While the results may not generalize to other ecosystems or CMSs, the methodology is transferable to ecosystems that provide library metadata such as e-mails or repository links. We focus on GitHub because prior work shows that PyPI libraries predominantly link to GitHub rather than to alternatives such as GitLab or Bitbucket~\cite{tsakpinis2024analyzing}.

\textbf{Reliability:} 
Although all data is collected from publicly accessible sources, reproducibility is limited by the dynamic nature of PyPI. Our dataset represents a snapshot at the time of collection, and since PyPI does not support historical queries, exact reproduction is not possible. To address this, we have published the dataset and analysis code on Figshare~\cite{tsakpinis_pretschner_2026}. Another threat arises from the continuous addition and modification of libraries, which may result in an outdated dataset. To mitigate this, the study should be repeated.

%% file: sections/7_Conclusion_and_Future_Work.tex
\section{Conclusion and Future Work}

This paper presents a large-scale empirical analysis on the availability of maintainer e-mail addresses across the PyPI ecosystem and its dependency structures. Such contact information plays a crucial role in enabling direct communication between maintainers and external stakeholders, particularly for sensitive or security related topics that require private channels. By analyzing 754,413 libraries, we show that a large majority (79.1\%) provide at least one valid e-mail address, with PyPI serving as the primary source of contact information. Availability increases with ecosystem importance and is particularly high within dependency chains, where up to 97.7\% of direct and 97.5\% of transitive dependencies provide contact information. At the same time, we identify many invalid entries, primarily due to missing e-mail fields, which may also reflect maintainers’ preferences to balance accessibility with privacy and security concerns. Overall, our results suggest that maintainer reachability in the PyPI ecosystem is generally strong, especially for widely reused libraries, while also highlighting opportunities to improve metadata completeness through lightweight interventions.

Looking ahead, future work should qualitatively explore the motivations behind maintainers’ decisions to include contact information, complementing and validating our assumptions about e-mail provision, as well as the barriers to providing such information to derive targeted interventions. In this context, we also plan to investigate why PyPI is the preferred location for listing e-mail addresses, while GitHub is often omitted. Finally, we will periodically replicate this study to track how the availability of contact information evolves over time.

%% file: sections/8_Data_Availability.tex
\section{Data Availability}
\label{sec:data_availabaility}

The dataset supporting this study, along with the analysis artifacts, is openly available on Figshare under a CC-BY 4.0 license~\cite{tsakpinis_pretschner_2026}. The replication package contains all materials necessary to reproduce the reported results. The only exception concerns the domain-level analyses used in the construct validity discussion: while the analysis code and aggregated results are provided, the raw e-mail addresses collected from PyPI and GitHub are excluded to protect the privacy of authors and maintainers across the PyPI ecosystem.